\documentstyle{article}
\begin{document}

\title{The structure of singularities in inhomogeneous cosmological models.}
\author{Alan D. Rendall, Max-Planck-Institut f\"ur Gravitationsphysik,
\\Schlaatzweg 1, 14473 Potsdam, Germany}

\maketitle

\begin{abstract}
Recent progress in understanding the structure of cosmological singularities
is reviewed. The well-known picture due to Belinskii, Khalatnikov and 
Lifschitz (BKL) is summarized briefly and it is discussed what existing 
analytical and numerical results have to tell us about the validity of this 
picture. If the BKL description is correct then most cosmological 
singularities are complicated. However there are some cases where it 
predicts simple singularities. These cases should be particularly 
amenable to mathematical investigation and the results in this direction
which have been achieved so far are described.
\end{abstract}

\section{Introduction}
The purpose of this article is to survey some recent progress in 
understanding the structure of cosmological singularities. Many years ago, 
Belinskii, Khalatnikov and Lifshitz (BKL) proposed a picture of singularities 
in general inhomogeneous spacetimes\cite{bkl82}. This was based on formal 
calculations which until recently had never been supported by either rigorous 
mathematical proofs or careful numerical calculations. This picture will be
referred to in the following as the BKL scenario.

A key element of the BKL scenario is that near a singularity the evolution
at different spatial points decouples. This of course requires a specification
of what is meant by a spatial point. As a first attempt at doing this we may 
suppose that a suitable local coordinate system is used in the definition. The 
decoupled evolution then corresponds to a spatially homogeneous model at
each spacetime point. A second element is that generically the matter content
of spacetime should have a negligible effect on the dynamics near the 
singularity. Thus the essentials of the dynamics in the generic case should
be described by vacuum spacetimes. This is why vacuum models are believed 
to be relevant to the study of cosmological singularities even although the
energy density blows up there. (This can still have a negligible dynamical
effect if other quantities in the Einstein equations blow up even faster.)
A third element is a description of the dynamics of the most general 
spatially homogeneous models near the singularity, which is supposed to be 
modelled by a spacetime of Bianchi type IX. Combining the vacuum condition
with the Bianchi type IX condition leads to the Mixmaster solution, which
thus takes on a central role in these considerations. A characteristic
feature of the Mixmaster solution is that it exhibits complicated
oscillatory behaviour.

It should be noted that the statement that the matter is generically 
negligible near the singularity is subject to certain restrictions.
For instance, within the BKL scenario a massless scalar field as matter
model remains important up to the singularity and kills Mixmaster
oscillations. As we will see later, an electromagnetic field can remain
important near the singularity, causing Mixmaster-like behaviour to 
appear in symmetry classes which show no Mixmaster behviour in vacuum.
There are well-known criteria for deciding which matter models can be
neglected and what will be their effect if they cannot but these will
not be discussed here. The role of Bianchi type IX is that it and Bianchi
type VIII, which has a very similar behaviour near the singularity, are
supposed to be the most general spatially homogeneous models. It seems, 
however, that there is a class of Bianchi type VI${}_h$ models with $h=-1/9$
which are equally general and their role in the BKL scenario remains to be 
clarified. (See \cite{wainwright97}, section 8.1, for a discussion of this.)
  
Up to now the BKL scenario is neither supported nor invalidated by rigorous
mathematical arguments. In fact even the dynamics of the Mixmaster solution, 
which serves as a basic model in the scheme, is not understood mathematically.
Some partial results, together with a description of some of the open 
questions can be found in \cite{rendall97a}. On the other hand, heuristic and
numerical work combine to give good evidence for the BKL scenario in the
spatially homogeneous case.

In the spatially inhomogeneous case numerical approaches had, until recently,
been no more successful than mathematical ones. The decoupling and pointwise
Mixmaster behaviour predicted by BKL had not been reliably seen numerically
in any spatially inhomogeneous case. There is an important reason for this
which will now be explained. One approach to looking for a simple example
would be to start with Mixmaster initial data and do a spatially 
inhomogeneous perturbation while keeping as much symmetry as possible, so as 
to make the problem as simple as possible. Unfortunately, the Mixmaster
model and its close relative of Bianchi type VIII have the property that 
no two linearly independent Killing vectors commute. Thus it is impossible 
to perturb away from three Killing vectors without going all the way down
to one Killing vector. The intermediate case of two Killing vectors is not
attainable. In this way one is led to a numerical problem where the effective
space dimension is two. It is very difficult to produce numerical results
which can be trusted in a context which combines Mixmaster-like behaviour
and two spatial dimensions.

A possible way round this difficulty was suggested in \cite{rendall97b}. This
arises from the fact that solutions of Bianchi type VI${}_0$ with 
electromagnetic fields show Mixmaster
behaviour (see \cite{leblanc95}) and that, since the Lie algebra which defines 
the symmetry of Bianchi type VI${}_0$ has a two-dimensional abelian
subalgebra, solutions of this type can be perturbed to solutions with two
Killing vectors. (The discussion in \cite{rendall97b} suffers from the 
misconception that matter is required in these models. Actually the 
source-free Einstein-Maxwell equations suffice.) In this way inhomogeneous
models are obtained where the effective space dimension is one. 

This possibility has been exploited recently by Weaver, Isenberg and Berger
\cite{weaver98}. They showed numerically that, in a class of solutions of the 
Einstein-Maxwell equations, independent Mixmaster-like oscillations are 
observed at different spatial points as the singularity is approached, 
agreeing with the BKL picture in that case. Thus for the first time a
solid confirmation of the applicability of the BKL scenario to a class of
inhomogeneous spacetimes showing Mixmaster behaviour has been obtained.

\section{Simple Singularities}

There are special situations in which the BKL picture predicts relatively
simple singularities and in this case one may hope to prove rigorous theorems
on their structure. Recently results of this kind have been proved using a 
class of singular partial differential equations, the Fuchsian equations. An 
introduction to the theory of these equations can be found in 
\cite{kichenas96}. The basic idea in applying these techniques to construct
singular solutions of partial differential equations is as follows. Write the 
solution $u$ as $u_0+u_1$, where $u_0$ is an explicit expression which is 
singular and $u_1$ is unknown but supposed to be regular. Now rewrite the 
original equation in terms of the unknown $u_1$. The resulting equation for 
$u_1$ will in general be singular, even if the original equation was regular.
In other words the task of finding singular solutions of a regular equation
is transformed into that of finding regular solutions of a singular equation.
In favourable cases the singular equation is of a special (Fuchsian) form
and it is possible to prove the existence of a unique solution for prescribed
singular part $u_0$. The end result is that one can solve a sort of Cauchy
problem with data on the singularity.

Fuchsian techniques can be used to give statements about the singularities
of general solutions in a particular class on different levels. In order of
increasing sophistication these are as follows. The first level is to prove
existence of solutions corresponding to analytic data on the singularity
which depend on the same number of free functions as the general solution.
The next level is the corresponding statement with smooth rather than analytic
data. The final one is to show that the solutions obtained in the second 
step include all those evolving from a non-empty open set of initial data
on a regular Cauchy surface. Proofs on all these levels are available in 
certain examples. 

One of the advantages of Fuchsian techniques is that they do not depend on
the spatial dimension. Thus it is possible to get results on hyperbolic 
equations in any space dimension which could not be achieved by the usual
approaches to hyperbolic equations available today. In general relativity
this means that results may be obtained on spacetimes with only one Killing
vector or none at all. Techniques available up to now for studying the 
structure of spacetimes determined by Cauchy data belonging to a certain
class near their singularities depended on having 
at least two Killing vectors so that the effective space dimension of the
system of partial differential equations was one.

Potential applications of these techniques in general relativity to cases
with little or no symmetry will now be listed. A case where this approach 
has been carried out successfully is that of isotropic singularities. The
case of a perfect fluid with radiation equation of state has been analysed
by Claudel\cite{claudel98} and generalizations to other fluids and to
kinetic theory have been obtained by Tod\cite{tod97} and 
Anguige\cite{anguige97}. Cases which have not yet been worked out are
those of the Einstein-scalar field system in four dimensions (quiescent 
cosmology), the Einstein vacuum equations in sufficiently high dimensions,
and vacuum solutions in four dimensions with polarized $U(1)$-symmetry. 

It may be asked why, if the Fuchsian approach is applicable without symmetry
assumptions, the above examples are difficult to handle. The reason is that
while the analytical theory is independent of symmetry, the algebra required
to reduce a particular example to Fuchsian form may be very complicated.

To conclude, some examples will be mentioned where simple singularities have
been successfully treated. The first is that of the Gowdy spacetimes. In
\cite{kichenas98} it was shown that there exist Gowdy spacetimes with
singularities of a very particular (velocity-dominated) form which depend
on the maximum number of free analytic functions. Recall that the Gowdy 
spacetimes are characterized by the fact that they are solutions of the vacuum 
Einstein equations, that they have two commuting spacelike Killing vectors,
that they are spatially compact and that the so-called twist constants vanish.
The spatial compactness is not relevant to the results of \cite{kichenas98}.
Later a similar result was obtained in the case where the condition on the
twist constants is dropped while the situation considered is specialized in
another direction (restriction to polarized models)\cite{isenberg98}

The conclusion is that first steps have been taken towards a rigorous
mathematical treatment of \lq simple\rq\ singularities with a number of
examples already having been worked out. Moreover, there is a potential
for significant progress in this area in the near future. Finally, it 
should be noted that the results presented in this section and the last
indicate the prospect of a fruitful marriage of numerical and analytical
techniques in the study of the structure of singularities in inhomogeneous
cosmological models.


\begin{thebibliography}{10}

\bibitem{anguige97} Anguige, K.: Isotropic cosmological singularities. DPhil
Thesis, University of Oxford, 1997.

\bibitem{bkl82} Belinskii, V. A., Khalatnikov, I. M. and Lifschitz, E. M.:
A general solution of the Einstein equations with a time singularity.
Adv. Phys. 13, 639-667 (1982).

\bibitem{claudel98} Claudel, C. M.: The Cauchy problem for quasi-linear 
hyperbolic evolution problems with a singularity in the time. Preprint,
to appear in Proc. R. Soc. Lond.

\bibitem{isenberg98} Isenberg, J. and Kichenassamy, S.: Asymptotic behaviour 
in polarized $T^2$-symmetric vacuum spacetimes. Max-Planck-Institut f\"ur
Mathematik in den Naturwissenschaften, Leipzig, Preprint No. 13.

\bibitem{kichenas96} Kichenassamy, S.: Nonlinear Wave Equations. Marcel
Dekker, New York, 1996.

\bibitem{kichenas98} Kichenassamy, S. and Rendall, A. D.: Analytic description 
of singularities in Gowdy spacetimes. Preprint, to appear in Class. Quantum 
Grav.

\bibitem{leblanc95} Leblanc, V. G., Kerr, D. and Wainwright, J.: Asymptotic
states of magnetic Bianchi VI${}_0$ cosmologies. Class. Quantum Grav.  12, 
513-541 (1995).

\bibitem{rendall97a} Rendall, A. D.: Global dynamics of the mixmaster model.
Class. Quantum Grav.  14, 2341-2356 (1997).

\bibitem{rendall97b} Rendall, A. D.: Solutions of the Einstein equations with
matter. In M. Francaviglia, G. Longhi, L. Lusanna, E. Sorace (eds.) 
Proceedings of the 14th International Conference on General Relativity and 
Gravitation, pp.313-335. World Scientific, 1997.

\bibitem{tod97} Tod, K. P.: Isotropic cosmological singularities. (to appear 
in the Proceedings of the XIIth International Congress of Mathematical Physics,
Brisbane, 1997.)

\bibitem{wainwright97} Wainwright, J. and Ellis, G. F. R. (eds.): Dynamical
systems in cosmology. Cambridge University Press, Cambridge, 1997.

\bibitem{weaver98} Weaver, M., Isenberg, J., and Berger, B. K.: Mixmaster 
behaviour in inhomogeneous cosmological models. Phys. Rev. Lett. 80, 
2984-2987 (1998).

\end{thebibliography}
\end{document}